\documentclass[aps,amssymb,amsmath,reprint]{revtex4-1} 

\usepackage{graphicx}  % needed for figures
\usepackage{dcolumn}   % needed for some tables
\usepackage{physics}
\usepackage{color}
\usepackage{appendix}
\usepackage[normalem]{ulem}
\usepackage{lineno}
\definecolor{orange}{rgb}{0.8, 0.3, 0}

\definecolor{blueviolet}{rgb}{0.2, 0.2, 0.6}
\usepackage[pdftex, % FOR PDFLATEX ONLY
bookmarks=true, % Bookmark bar
colorlinks=true,
allcolors=blueviolet,
pdfstartview={FitH}, % FitBH
]{hyperref}

\usepackage{comment}

\begin{document}

\title{In-situ scanning gate imaging of individual two-level material defects in live superconducting quantum circuits}
\author{M. Hegedüs$^{1,2,\ddagger}$}
\author{R. Banerjee$^{1,\ddagger}$}
\email{riju.banerjee@npl.co.uk}
\author{A. Hutcheson$^{1}$}
\author{T. Barker$^{1}$}
\author{S. Mahashabde$^3$}
\author{A. V.~Danilov$^3$}
\author{S. E.~Kubatkin$^3$}
\author{V. Antonov$^{2}$}
\author{S. E. de Graaf$^{1}$}
\email{sdg@npl.co.uk}

\affiliation{$^1$National Physical Laboratory, Teddington TW11 0LW, United Kingdom}
\affiliation{$^{2}$Physics Department, Royal Holloway University of London, Egham, United Kingdom } 
\affiliation{$^3$ Department of Microtechnology and Nanoscience MC2, Chalmers University of Technology, SE-412 96 G\"oteborg, Sweden}
\affiliation{$^\ddagger$ These authors contributed equally.}

%% Abstract #######################################
\begin{abstract}
The low temperature physics of structurally amorphous materials is governed by two-level system defects (TLS), the exact origin and nature of which remain elusive despite decades of study. Recent advances towards realising stable high-coherence platforms for quantum computing has increased the importance of studying TLS in solid-state quantum circuits, as they are a persistent source of decoherence and instability. Here we perform scanning gate microscopy on a live superconducting quantum circuit at millikelvin temperatures to locate individual TLS. Our method directly reveals the microscopic nature of TLS and is also capable of deducing the three dimensional orientation of individual TLS electric dipole moments.  Such insights, when combined with structural information of the underlying materials, can help unravel the detailed microscopic nature and chemical origin of TLS, directing strategies for their eventual mitigation. 
\end{abstract}
\maketitle

%\section{Introduction}
Glassy disordered materials show remarkable universality in their low-temperature thermal, acoustic and microwave absorption properties, irrespective of their chemical compositions \cite{RevModPhys.83.587, Leggett2013}. To explain these intriguing observations, Anderson et. al. and Phillips proposed the microscopic model of these materials to be dominated by low energy imperfections, namely, two-level system defects (TLS) \cite{Anderson1972, Phillips1972}, with their collective behaviour described by the phenomenological Standard Tunnelling Model (STM) \cite{phillips1987two}. However, in the half century since, direct probing of individual TLS defects and testing the microscopic principles underlying these macroscopic observations have been very challenging, fuelling intense debates over the exact nature of these defects \cite{muller2019towards, Leggett2013}. 

Though TLS have been extensively studied in glassy materials, recent advances in quantum computation and sensing have further underscored the need to characterise their chemical and structural properties \cite{muller2019towards, degraaf2022}. TLS are a major source of noise and decoherence, and even a single defect can spoil the performance of an entire circuit \cite{kilmov2018, muller2019towards, PhysRevResearch.6.033175}. Achieving fault tolerant quantum computing requires stable, high coherence qubits, which in turn need new tools to find, characterise, and understand the nature of TLS defects as they appear in live circuits. This task has been particularly challenging, as the low energy scales of the defects render them inaccessible to much of conventional material science techniques \cite{degraaf2022}. 

Established in-operando methods of detecting TLS, for example by tuning a qubit into resonance with it \cite{PhysRevLett.93.077003, muller2019towards}, or similarly tuning TLS by applying strain \cite{gabowskij2012} or electric \cite{degraaf2021, lisenfeld2015, bilmes2019, bilmes2020, bilmes2021quantum, lisenfeld2016, lisenfeld2019, degraaf2022} fields, are incapable of directly extracting precise defect locations. Furthermore, they only indirectly suggest the microscopic origin, revealing no chemical or structural information. On the other hand, traditional scanning probe techniques such as scanning tunnelling microscopy or atomic force microscopy (AFM) are routinely used for defect characterisation and offer atomic scale spatial resolution. Unfortunately, these techniques fall short by orders of magnitude in resolving the typical TLS energy scale. Despite this, scanning probe imaging of quantum circuits is becoming increasingly important for characterising electromagnetic field distributions and material properties in  quantum devices at low temperatures \cite{Lang2004Pinhole, geaney2019, degraaf2013, zhang2024ultrabroadband, oh2021, denisov2022, Marchiori2022}.

Here we integrate scanning gate microscopy (SGM) with in-situ readout of live superconducting quantum circuits at millikelvin temperatures to locate individual TLS defects, directly demonstrating their microscopic nature. The SGM can also operate in AFM mode for imaging device topography. Furthermore, we also deduce the electric dipole moment orientation of individual TLS, information that has remained elusive since they were conceptually put forward over half a century ago. We posit that by combining information about TLS orientation with detailed material structure and ab-initio calculations, our approach could help identify the origin and physical nature of these defects, leading to better solid state quantum circuits.

\begin{figure*}
\centering
\includegraphics[width=0.98\textwidth]{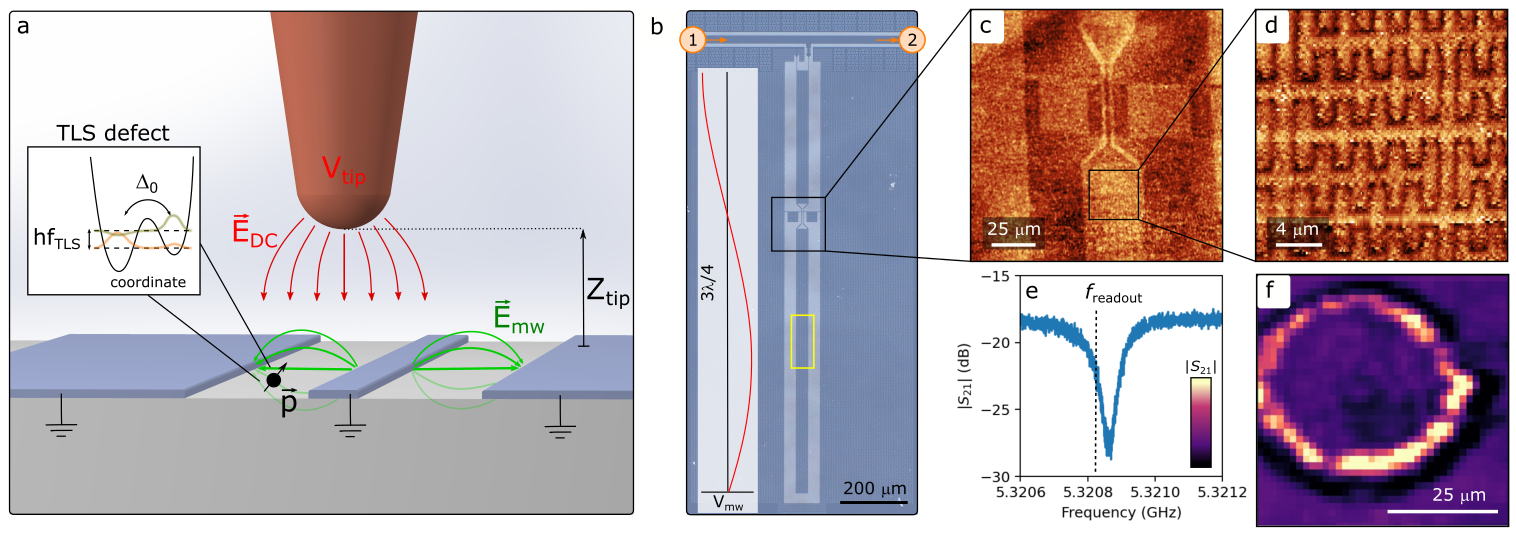}
\caption{\label{fig:q} {\bf{Experimental setup for imaging TLS}} (a) A schematic of our setup shows a sharp tip above a live circuit in which TLS defects reside. The TLS dipole moment $\vec{p}$ couples to the circuit via the microwave electric field $\vec{E}_{\rm mw}$. The tip is used for both AFM imaging and applying localised electric fields $\vec{E}_{\rm DC}$. (b) Optical image of the $3\lambda/4$ superconducting hanger resonator patterned in 40 nm thick NbN film on sapphire used in our study. The yellow rectangle indicates the location of the scan presented in Fig. 2. The inset shows the microwave voltage amplitude along the resonator. (c) Wide-area AFM scan over the live circuit at 200 mK, taken in the region indicated by the black box in (b). (d) Higher resolution AFM image shows the interdigitated capacitors within the resonator. (e) Example of the microwave transmission $S_{21}(f)$, measured between the ports marked 1 and 2 in (b), around the circuit resonance frequency $f_{\rm res}$. To detect TLS while scanning, we use a heterodyne readout scheme that measures the transmission at $f_{\rm readout}$, slightly detuned from $f_{\rm res}$. (f) Detecting a TLS. The $S_{21}(f_{\rm readout})$ transmission signal (color scale) as a function of tip position reveals a bright contour corresponding to a constant electric field from the tip at the location of a TLS in the center. The data was taken with $Z_{\rm tip}=15$ $\mu$m and $V_{\rm tip}=-10.25$ $V$. See text for further details.}
\end{figure*}

%\section{Experimental setup}
Our experimental setup combining SGM with in-situ device readout is described in the schematic presented in Fig. 1a. We use an electrochemically etched tungsten tip attached to a quartz tuning fork to facilitate AFM imaging. The tip is also connected to a voltage source for applying local electric fields in order to tune the energy of TLS. The entire setup is enclosed within a light-tight, magnetically shielded volume and suspended on springs below the mixing chamber plate of a dry dilution refrigerator to minimize vibrations. With this setup, we achieve a sample stage temperature of $\sim 45$ mK, as measured using a calibrated $\rm RuO_x$ thermometer.

Although our SGM setup is entirely agnostic to the circuit under study and is adaptable for examining any quantum component coupled to TLS, here we choose to study a superconducting resonator. Our sample consists of a $3\lambda/4$ resonator patterned in 40 nm thick NbN on sapphire, an optical image of which is shown in Fig. 1b.  Interdigitated capacitors concentrate the electric fields inside the resonator, enabling enhanced coupling to a large number of TLS defects. Further details about the sample can be found in Refs \cite{mahashabde_fast_2020, ranjan2022}. 

Figures 1c and 1d show AFM images of the live resonator obtained at 200 mK. This in-situ AFM imaging enables us to locate various device features. In all our experiments, we limit the scan speed to reduce heating of the piezo positioners and keep the sample temperature below 300 mK at all times, avoiding TLS saturation occurring for $k_BT>2hf_{\rm res}$ \cite{PhysRevB.80.132501} or TLS bath reconfiguration \cite{degraaf2020}. 

Following AFM imaging we locate individual TLS as follows. We position the tip at a constant height $Z_{\rm tip}$ above the sample, and at each grid point in the xy-plane conduct a tip voltage sweep. We continuously, in operation, monitor the microwave (MW) signal transmitted at a fixed frequency $f_{\rm readout}$ slightly offset from the resonance frequency $f_{\rm res}$ of the imaged device using a heterodyne detection measurement scheme (see Supplementary Information for details). A change in the measured signal $S_{21}(f_{\rm readout})$ thus means that either the resonator's centre frequency or quality factor has changed, the result of a TLS becoming resonant with the resonator. The microwave power level used for transmission measurements was kept very low to avoid saturating TLS, typically with an average photon population of 10-1000 in the resonator.  A slice of such a dataset $S_{21} (x, y, V_{tip}=\rm const)$ is shown in Fig. 1f, where data from a $40 \times 40$ 
 pixel grid is presented. The observed ring indicates a locus of points at which the tip needs to be positioned for the TLS to experience the same electric field magnitude, to bring it into resonance with the resonator, i.e. the TLS is located at the centre of the ring. In the Supplementary Information, we show more data taken at high microwave powers that confirms the saturation of the detected TLS.

\begin{figure*}
\centering
\includegraphics[width=0.85\textwidth]{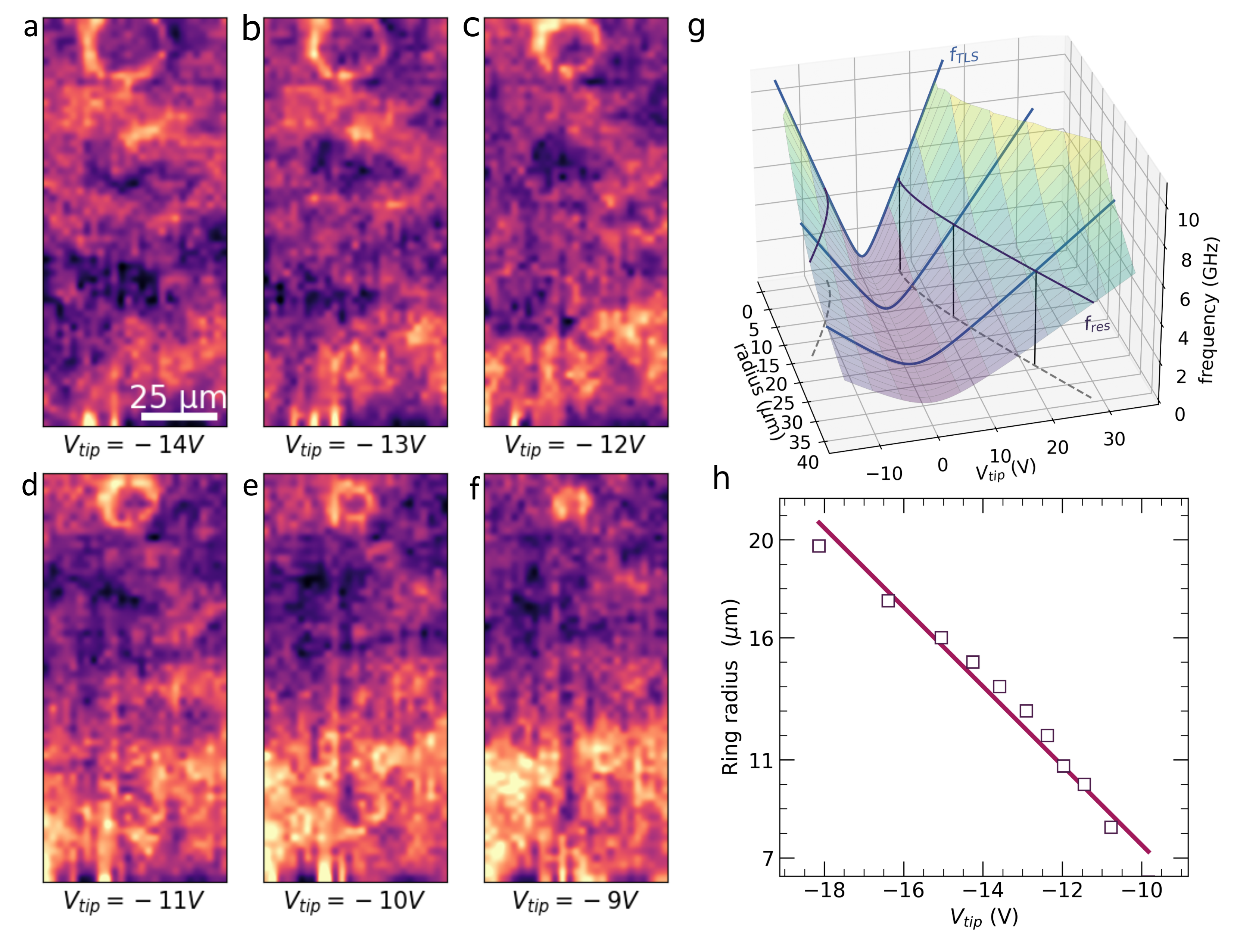}
\caption{\label{fig:q2} {\bf{Tuning TLS with scanning gate microscopy}} (a)-(f) Voltage slices taken with the tip $20 \ {\rm\mu}$m  above the sample show a bright, circular TLS ring near the top, along with local fluctuations in the plotted $S_{21}$ signal. Changing $V_{\rm tip}$ changes the size of the circular contour. The scans are taken in a $27 \times 60$ pixel grid over a $67.5 \ \mu$m $\times 150 \ \mu$m region marked by the yellow box in Fig. 1b. The scale bar in (a) is shared across the panels (a)-(f). The color scale shows the change in measured signal $S_{21}$ in arbitrary units. (g) A conceptual sketch showing the relationship between tip voltage and ring radius for a resonant TLS. Sweeping the tip voltage alters the TLS energy, according to eq. (\ref{eq:tls_hyperbola}) (blue curves). TLS hyperbolas intersect $f_{res}$ (black curves), at which point the TLS is detected by the resonator as a change in transmission $S_{21}$. (h) The radius of the ring in panels (a)-(f) (markers) decreases with increasing tip voltage. The solid line is a linear fit.}
\end{figure*}

%\section{Measurements}
Varying the voltage at which the two-dimensional slice is taken changes the ring diameter, as shown in Fig. 2a-2f (for a different TLS than in Fig. 1f). This data was taken at $Z_{\rm tip}=20 \ \rm\mu$m in the area of the yellow box in Fig. 1b.  In addition to the clear circular contour, we also note the fluctuating background arising from other nearby TLS outside the grid frame. The grid data was collected over three days, implying the observed ring and other experimental parameters remained stable throughout. All panels of Fig. 2a-2f have had a parabolic background subtracted to remove the capacitive contribution of the tip on the resonator, which typically gives a much larger (but independent of tip voltage) response than TLS. Reversing the tip voltage sweep direction does not affect the shape or location of the ring, ruling out any charging effects on the device. 

To understand why TLS manifest as rings in our experiment, we consider a TLS with an electric dipole moment $\vec{p}$ interacting with the microwave field $\vec{E}_{\rm mw}$ of a galvanically grounded resonator resulting in a coupling strength $hg = \vec{p}\cdot \vec{E}_{\rm mw}$. When a sharp AFM tip with a DC voltage $V_{tip}$ is positioned at a distance $r$ from the TLS, the defect experiences an electric field $\vec{E}_{\rm DC} \approx V_{tip}/r$, causing a shift in its energy level transition frequency
\begin{equation}
    hf_{\rm TLS} = \sqrt{\Delta_0^2 + (\epsilon + 2\vec{p}\cdot \vec{E}_{\rm DC})^2},
    \label{eq:tls_hyperbola}
\end{equation}
where $\Delta_0$ is the TLS asymmetry energy and $\epsilon$ is an offset energy imposed by the TLS's local environment \cite{lisenfeld2019}. As the tip voltage is swept, the TLS frequency is tuned along this hyperbolic trajectory (plotted as blue lines in Fig. 2g). When the TLS becomes resonant with the resonator frequency $f_{\rm res}$ (black curves in Fig 2g), it can be detected as a change in the measured $S_{21}$ signal. The contours traced out when moving the tip in the xy-plane are a set of points where the TLS experiences the same electric field magnitude.  For the ring shown in Fig. 2, a smaller (larger) tip voltage changes the distance $r$ at which the tip must be placed for the TLS's frequency to be shifted to be resonant with the resonator, and this results in a smaller (larger) ring. This shrinking of the ring with increasing tip voltage is also demonstrated in Fig. 2h, where we show that the radius of the ring in Fig. 2a-2f decreases linearly with the applied tip voltage for ring radii $\gtrsim Z_{\rm tip}$, as expected from Eq. (1).

\begin{figure*}
\centering
\includegraphics[width=0.8\textwidth]{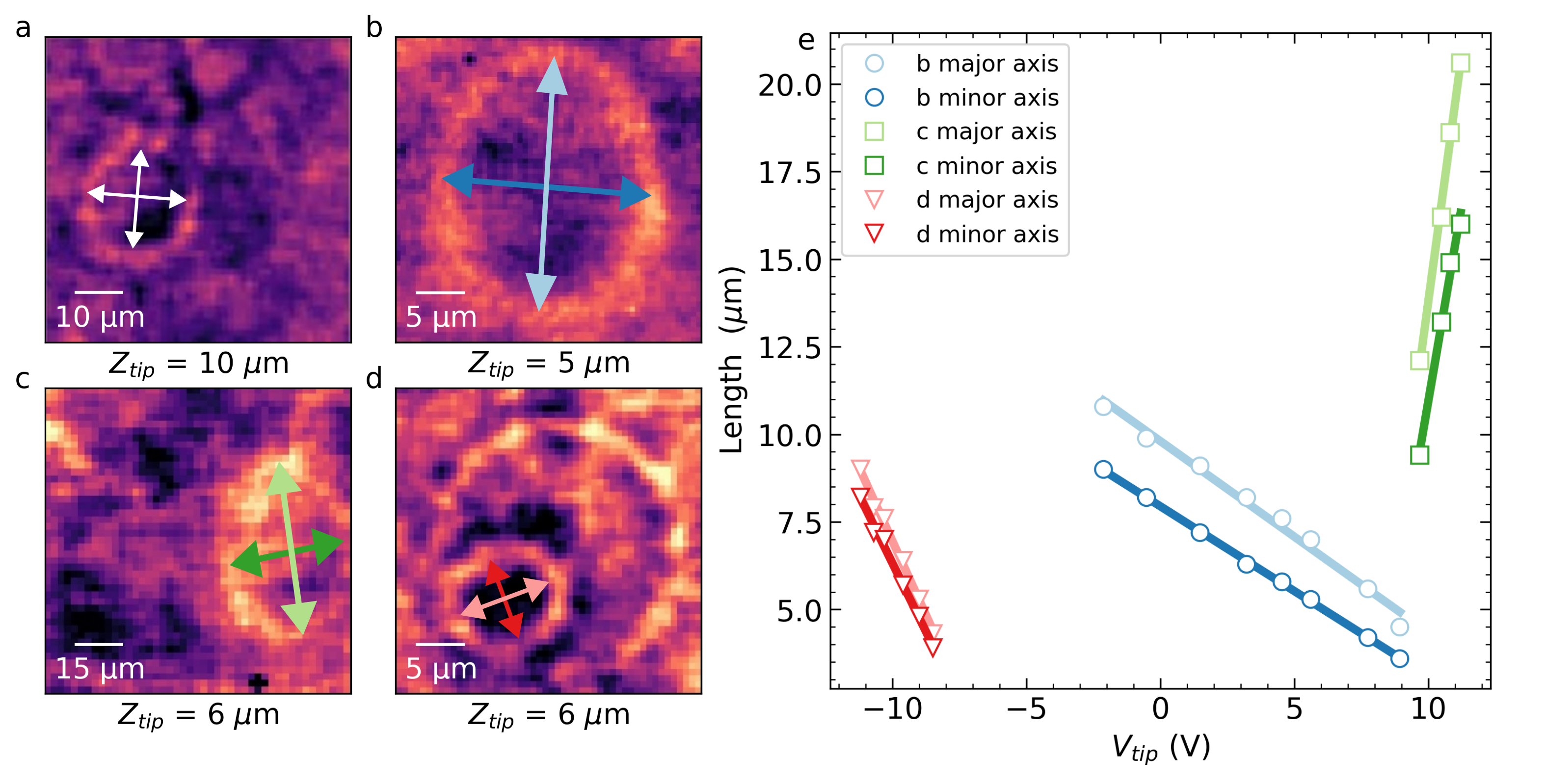}
\caption{{\bf{Zooming in on TLS}} (a) and (b) show close-up images of the TLS ring shown in Fig. 2, taken at (a) $Z_{\rm tip} = 10 \ \mu$m and (b) $Z_{\rm tip} = 5 \ \mu$m. (c) and (d) show two other TLS found elsewhere in the sample, and both images are with $Z_{\rm tip} = 6 \ \mu$m. Note the elliptical shape of the contours in (a)-(d), whose minor and major axes ($r_a$ and $r_b$ respectively) are marked by double-sided arrows. The contours appear circular for TLS dipole moments pointing mostly perpendicular to the sample surface, such as the large ring in (d). (e) Analogous to Fig. 2h, the major and minor axes of all three ellipses shrink or grow linearly with applied tip voltage. The straight lines are linear fits to the data. From the linear fits, we extract their minor/major ($r_a/r_b$) radius aspect ratios to be $r_a/r_b=0.95$ for (a), 0.8 for (b), 0.8 for (c) and 0.9 for (d). }
\end{figure*}

As the tip voltage is increased, in most cases, we expect to observe either a shrinking or a growing ring. For very large tip voltages one could expect to see two concentric contours originating from each of the two intersection points of $f_{\rm res}$ with the TLS hyperbola (Fig. 2h). These could be either both shrinking or both growing with increasing tip voltage. A third possibility is that one contour is shrinking, followed by one that is growing. These three scenarios depend on at what electric field strength with respect to zero field the TLS minima is located.
In the experiment, we kept $|\vec{E}_{\rm DC}|$  below $2 \times 10^6 \ V/m$ to prevent damaging the tip or sample, and hence, so far have only observed either shrinking or growing single rings.

Zooming in closer to the defect manifesting as the ring in Fig. 2 and bringing the tip closer to the sample surface results in the images in Fig. 3a and 3b with $Z_{\rm tip} =10 \ \mu m$ and $Z_{\rm tip}=5 \ \mu m$ respectively. Notably, when the tip is closer to the surface, the shape transforms from a circular to an elongated contour, which fits well to an ellipse. The major and minor axes of the fitted ellipse both shrink linearly with applied tip voltage as shown in Fig. 3e, and their ratio remains constant within experimental error. Zoomed-in images of a couple of other TLS defects that appear elliptic (located outside the field of view of Fig. 2) are shown in Fig. 3c and 3d for $Z_{\rm tip}\approx6 \ \mu$m. The in-plane orientations of all these ellipses differ and do not appear to correlate with any patterned device features. Instead, the elliptic shape is an indication of the TLS dipole moment orientation.

%\section{Simulations}
To extract this orientation, we proceed with modelling the expected response. Following \cite{CavitySpinCoupling2010}, the transmission $S_{21}(f)$ of the resonator when the tip voltage tunes a TLS to be resonant with it (i.e., for $f_{\rm TLS}\simeq f_{\rm res}$) can be expressed as 
\begin{equation}
    |S_{21}(f)| = \bigg|1 + \frac{\kappa_c}{i(f-f_{\rm res}) -\kappa + \frac{g^2}{i(f_{\rm TLS}-f_{\rm res})-\gamma_{\rm TLS}/2}}\bigg|,\label{eq:s21f}
\end{equation}
where $\gamma_{\rm TLS}$ is the TLS linewidth and $\kappa = \kappa_c + \kappa_i$ is the total resonator loss rate, given by the sum of the coupling and internal loss rates respectively. When the TLS-resonator coupling $g$ is weak ($\lesssim 100 $ kHz), the resonator response is mainly dissipative. In contrast, a large $g$ also induces shifts in the resonator frequency. Our simulations indicate that, experimentally, we more frequently encounter the former regime than the latter (an example of which is shown in Fig. 1f).

\begin{figure*}
\centering
\includegraphics[width=0.98\textwidth]{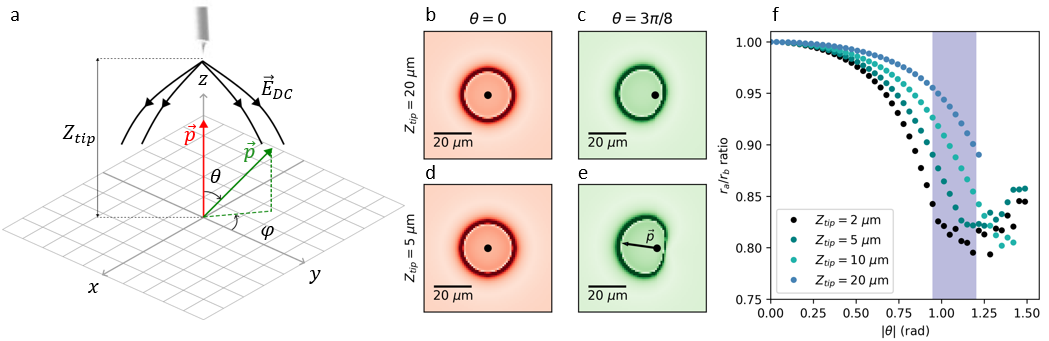}
\caption{\label{fig:sims}{\bf{Extracting TLS dipole moment orientations from simulations}} (a) Schematic of the system used for simulations. (b-e) Simulated SGM images for TLS on the substrate surface with two different polar angles and two different tip heights using a simplified geometry (see Supplementary information for details). When the dipole moment is oriented perpendicular to the sample plane (red arrow in (a)), the observed rings are always circular. The rings become elliptical when the TLS dipole is oriented to have a large in-plane component (green arrow in (a)) and the tip is brought close to the surface. The black dots indicate the location of the TLS, and the arrow in (e) indicates the simulated orientation of the in-plane dipole moment. (f) Simulations showing how the aspect ratio of ellipses varies with the tip-sample distance and polar angle $\theta$. For each data point, the tip voltage has been adjusted such as to produce contours with the same $r_b \approx 18.5$ $\mu$m, a scale close to the observed ellipses in Fig 3. For all the simulations we use experimentally consistent parameters of $\gamma_{\rm TLS}= 150 \ {\rm MHz}, g= 100 \rm kHz$, $f_{\rm readout}-f_{\rm res} = 20$ kHz and $\kappa = 100$ kHz.}
\end{figure*}

Information about the orientation of the TLS dipole moment is implicitly present in Eq. (\ref{eq:s21f}) via  $f_{\rm TLS}$, defined in Eq. (\ref{eq:tls_hyperbola}). Here the term $\Vec{p}\cdot \vec{E}_{\rm DC}$ implies that we can be selectively sensitive to the in-plane or out-of-plane component of $\vec{p}$ 
 by changing the direction of the applied electric field $\vec{E}_{\rm DC}$, enabled by a sharp tip. In Fig. 4a we define the TLS orientation by the in-plane ($\phi$) and out-of-plane ($\theta$) angles with respect to the tip coordinate system. 
 
 For a tip with a set voltage held at a particular height $Z_{\rm tip}$ above the sample, the electric field distribution $\vec{E}_{\rm DC}$ on the sample surface is computed in COMSOL. For a chosen orientation of the TLS dipole moment $\vec{p}$, the spatial $\vec{E}_{\rm DC}$ data is then used to calculate $S_{21}$ from Eq. (\ref{eq:s21f}).
 In Fig. 4b, for illustration, we show such simulated TLS images for two different orientations of a TLS dipole moment for two different $Z_{\rm tip}$. This shows the expected circular contour for $\theta = 0$ for both tip heights as the z-component of the electric field dominates in both cases. However, for large $\theta$ and small $Z_{\rm tip}$ the in-plane component of the electric field becomes increasingly important, and an elliptical contour emerges. From this elliptical contour, the in-plane angle $\phi$ of the TLS dipole moment can be inferred directly from the angle of the short axis of the contour, meaning we can directly read $\phi$ from the panels in Fig. 3a-3d.

Deducing the polar angle $\theta$ requires more detailed simulations.  We fit the simulated two-dimensional $S_{21}$ map (similar to images in Fig. 4c and 4e) to elliptical contours to extract the major ($r_a$) and minor ($r_b$) axes of the ellipses. The process is repeated as a function of $\theta$ and $Z_{\rm tip}$ to obtain Fig. 4f. The plot shows that the change in $Z_{\rm tip}$ from 10 to 5 $\mu$m and resulting contour aspect ratio experimentally seen in Fig. 3a and 3b, corresponds to a narrow (shaded) range of possible $\theta$ for this TLS. Also noteworthy is that the smallest possible $r_a/r_b$ ratio obtainable is 0.8 for large $\theta$, in good agreement with experimental observations as we have not seen any TLS with a smaller ratio. We note that in Fig. 4f, for very large $\theta$, the $r_a/r_b$ ratio again starts to increase. This is due to the resonance contour starting to deviate from an elliptical shape. 

The simulations indicate that for large $\theta$ the actual defect is located slightly off-center along the minor axis of the ellipse. Furthermore, they suggest that both the center of the ellipse and its aspect ratio change slightly with the applied tip voltage. Unfortunately, the large fields of view required to visualise these transitions are beyond the scan range of our microscope.

Although the slopes in Fig. 2h and 3e indicate the relative magnitudes of the TLS dipole moments,  uncertainty in the exact electric field at the defect, and the unknown $\epsilon$ of each defect prevents extraction of $|\vec{p}|$. However, to reproduce the experimental results in our modelling, we assume $|\vec{p}|=1$ e\AA~\cite{lisenfeld2019, degraaf2018, PhysRevLett.116.167002} which results in tip voltages very close to those used in our experiments. To precisely extract $|p|$ simultaneous frequency tuning of the device is required \cite{lisenfeld2019}. 

%\section{Discussions and Outlook}
Of the multitude of TLS present in the sample, our experimental setup is most sensitive to only those defects that are near the surface, close to the resonator and have stronger coupling to it. Thus, the observed TLS ought to be those most debilitating to device coherence.

Several factors can affect the shape of the observed contours. The width of the contour is related to the quality factor of the resonator, the linewidth of TLS, and $\partial f_{\rm TLS}/\partial r$. 
The spatial resolution is also limited by the sharpness of the AFM tip. Scanning electron microscope images of the tip, captured before and after the experiment (see Supplementary Information for details), demonstrate that the tip apex was never larger than a few microns during the six month experiment, setting our resolution. Out-of-plane tilting of the sample can also distort the contours. However, the $\sim 2^{\rm o}$ tilt of the sample measured using AFM should result in a tip-sample height difference of only $\approx 700 \ \rm nm$ over a $20 \ \rm \mu m$ scan range, resulting in minimal distortion.

Another factor blurring the experimental images at small $Z_{\rm tip}$ is the mechanical vibrations of the tip. Taking grid measurements with very small tip-sample distances were difficult, likely due to these mechanical vibrations in our system coming from the pulse tube cryocooler. In particular,  attempting a grid over the defect in Fig. 3b at $Z_{\rm tip} = 1$ $\mu$m resulted in the TLS ring disappearing also in subsequent scans at larger $Z_{\rm tip}$. We speculate this to be because of accidental contact with the defect, potentially also demonstrating the delicate glassy state of the TLS, susceptible to even minute perturbations.  
More stable scanning will allow closer imaging and improved resolution down to 10's of nm, ultimately limited by the achievable in-plane electric field gradient from the tip, set by the tip size.

Future experiments will undoubtedly have better control over these aspects, leading to a more precise determination of $\theta$. Importantly, our simulations show that as long as the TLS defect is located a sufficiently large distance from any metalisation on the sample/device, the specific device geometry does not play a significant role. For TLS located very close to metallic structures on the sample, however, the electric fields will always be perpendicular to the metal, and hence these TLS will always appear as circular contours. It also means that the TLS observed here as ellipses are located on or in the dielectric substrate.

Several improvements to our experiment can yield significantly more information about the defects. For example, utilising a substrate back gate and applying an additional variable electric field in the vertical direction can pinpoint the location of defects in three dimensions. Recent studies have attributed the majority of decoherence in superconducting devices to surface losses \cite{PhysRevApplied.21.044021, bilmes2020, 10.1063/1.3637047}, and a systematic study of TLS concentrations with height would be very beneficial to this discussion.

The bulk of our knowledge about TLS in superconducting quantum circuits is based on phenomenological models deduced from observing the behaviour of the devices they inhabit. Direct interrogation of individual defects is rare, and due to the complexity of the setup required, relatively little attention has been focused on understanding the physical and chemical nature of these defects. Our approach of combining scanning probe systems with live quantum circuit readout is a promising direction for further understanding decoherence mechanisms, and testing the validity of the standard tunneling model. Localising the defects using our technique will also facilitate their study using other established scanning probe and surface analysis techniques. Studying multiple devices and different fabrication techniques can help generate statistics on the concentration and origins of TLS in different materials. Such experimental approaches, coupled with atomistic modeling \cite{PhysRevLett.111.065901, PhysRevB.98.020403, un2021, Cyster2021} can aid in the understanding and eventual mitigation of TLS defects.

%\section*{Data availability}
\section*{Acknowledgements}
 We thank A. Tzalenchuk and T. Lindstrom for helpful discussions. We acknowledge the support from the UK Department for Science, Innovation and Technology through the National Measurement System (NMS), the Engineering and Physical Sciences Research Council (EPSRC) (Grant Number EP/W027526/1), and Google Faculty research awards. S.K. and A.D. acknowledge the support from the Swedish Research Council (VR) (Grant Agreements No. 2019-05480 and No. 2020-04393).

%\section*{Author contributions statement}

%\section*{Competing interests statement}

\bibliography{main}

\clearpage
\newpage
\setcounter{figure}{0}
\renewcommand{\figurename}{{{Supplementary FIG.}}}
\renewcommand{\tablename}{{{Supplementary Table}}}
\onecolumngrid
\begin{centering}{\large{\bf{SUPPLEMENTARY INFORMATION}\\}}\end{centering}\vspace{18mm}
\twocolumngrid
%%%%%%%%
\section{Superconducting resonator sample}
The Nb on Sapphire resonator sample used in this work consists of a $3 \lambda/4$ hanger-type resonator. It consist of two parallel prongs which serve as inductors, connected together on one side of the resonator. A series of interdigitated capacitors couple the two prongs to each other. The interdigitated capacitors concentrate the electric fields inside the resonator, thereby enabling stronger coupling to TLS, facilitating easier detection. The two prongs are connected to the ground plane on either side of the resonator (through an inductive filter, seen in the scan of Fig. 1c), to facilitate the application of a DC current through the resonator. This can be used to tune the resonance frequency, a property not utilised in this work. 

The $3 \lambda/4$ fundamental mode of the resonator results in the microwave voltage standing wave amplitude as sketched in Fig. 1b. The inductive filters are located in the voltage node to maximise the quality factor of the resonator. 
In this region the sensitivity to TLS is reduced as here they couple weakly to the resonator microwave field. Instead we detect the TLS most detrimental to device performance near the voltage anti-nodes.

\section{Experimental microwave setup}

\begin{figure}[h!]
\centering
\includegraphics[width=0.44\textwidth]{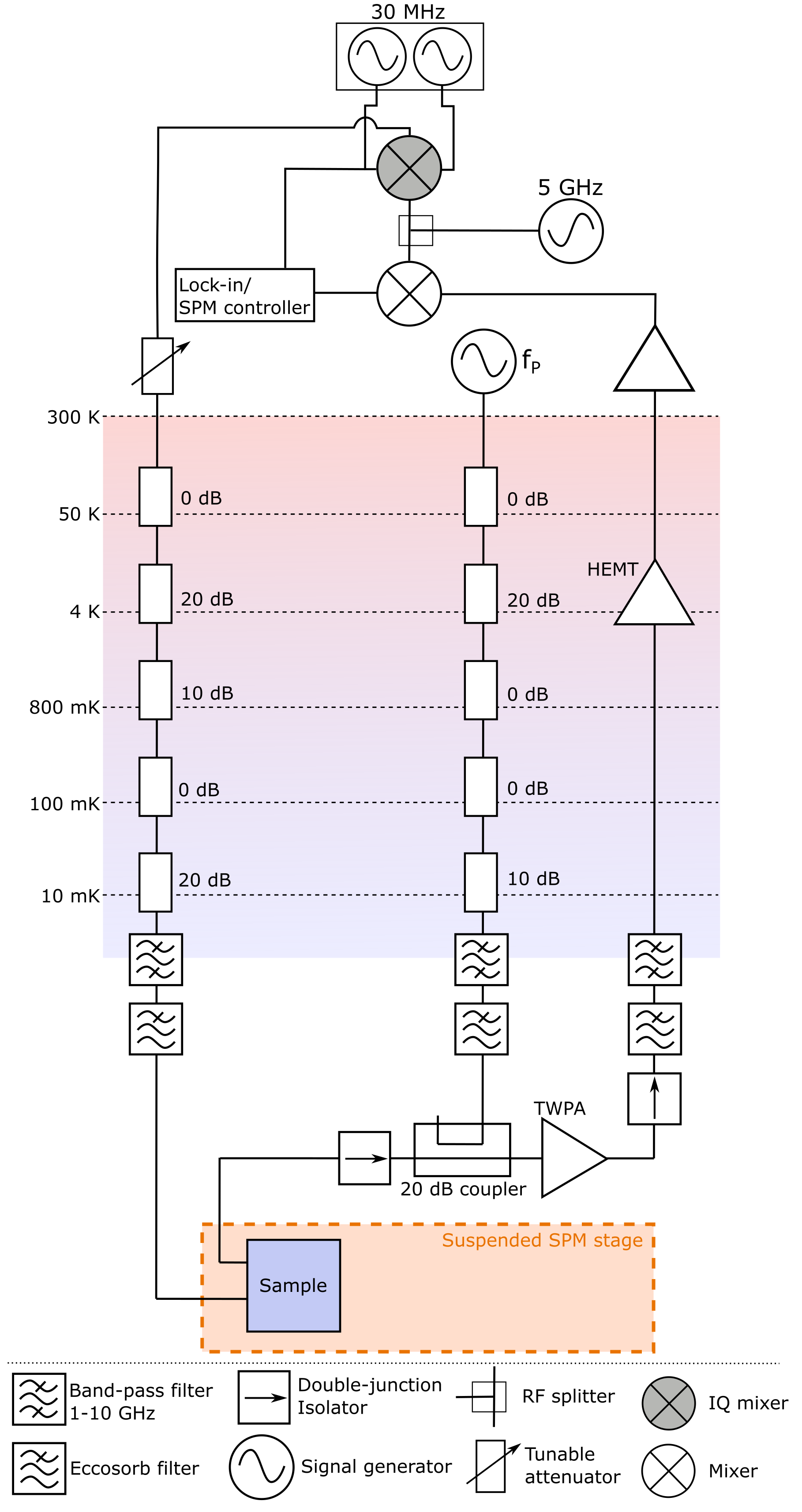}
\caption{\label{SI_fig:RFsetup} Heterodyne measurement setup and fridge RF wiring. }
\end{figure}

Supplementary Fig. \ref{SI_fig:RFsetup} shows the configuration of the microwave wiring inside the dilution refrigerator, and the setup of the heterodyne detection scheme at room temperature. In brief, two low-frequency (30 MHz) phase-shifted signals are up-converted to a single side-band tone at the sample resonance frequency, and passed down a heavily attenuated coaxial line to the sample on the SPM stage. The signal from the sample is then returned via a travelling-wave parametric amplifier (TWPA, SilentWaves Argo; driven by a pump tone at $f_P\approx 6$ GHz) before being further amplified by a high-electron mobility transistor (HEMT) amplifier at the 4K stage of the cryostat, and further amplified at room temperature to the desired level. The signal is then again down-converted to 30 MHz and demodulated using a lockin-amplifier, which feeds the two analog demodulated quadrature signals, proportional to the microwave transmission at the chosen frequency $S_{21}(f)$, to the SPM control electronics (Nanonis). We record the data from both quadratures, but rotate the phase such as to put most of the signal in one of the quadratures. For simplicity, the data shown in the manuscript is from one of these quadratures only.

\section{SEM images of the tip}
The tip used for AFM and applying the local gate voltage was produced by etching a 0.25 mm tungsten wire in a KNO solution. The etched tip was cleaned in deionised water to both stop the etching process and clean any residual salts sticking to it. The tip was then imaged using SEM to verify its sharpness. An SEM image of the tip taken before scanning is shown in Supplementary Fig. \ref{SI_fig:tip}a. 

Images taken after scanning for six months (Supplementary Fig. \ref{SI_fig:tip} b and c) show that it became blunt over time. Nevertheless, the end diameter was still less than $5 \ \mu m$. The simulation results presented in this work used a tip with a diameter of $\lesssim 5 \ \mu m$ to accurately model our observations.

\begin{figure}
\centering
\includegraphics[width=0.32\textwidth]{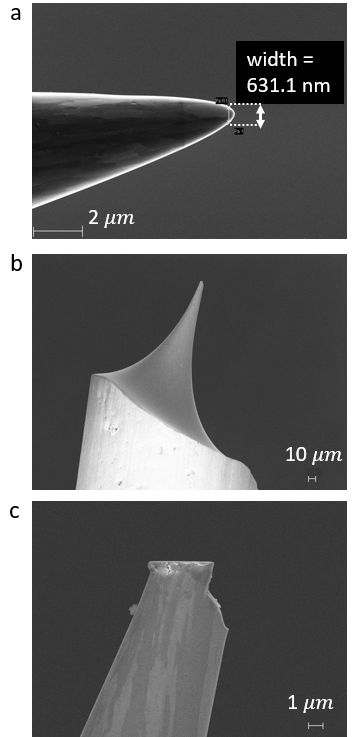}
\caption{\label{SI_fig:tip} \textbf{SEM images of the tip.} (a) SEM image of the tip taken before scanning shows a sharp (sub-micron) tip. Large area (b) and zoomed-in (c) SEM images of the tip taken after scanning. The end is blunted due to scanning for 6 months, but is still less than 5 microns wide.}
\end{figure}

\section{TLS saturation and power dependence}
A common signature of TLS is through their power dependence. At low microwave powers, TLS can absorb photons from the resonator, making TLS the primary source of loss in the circuit. At higher microwave powers, TLS cannot dissipate the absorbed energy as phonons quickly enough, causing them to saturate. In Supplementary Fig. \ref{SI_fig:power-Q} we show the power dependence of the internal and external (coupling) quality factors of the resonator in which TLS were imaged in this work. 

\begin{figure}
\centering
\includegraphics[width=0.48\textwidth]{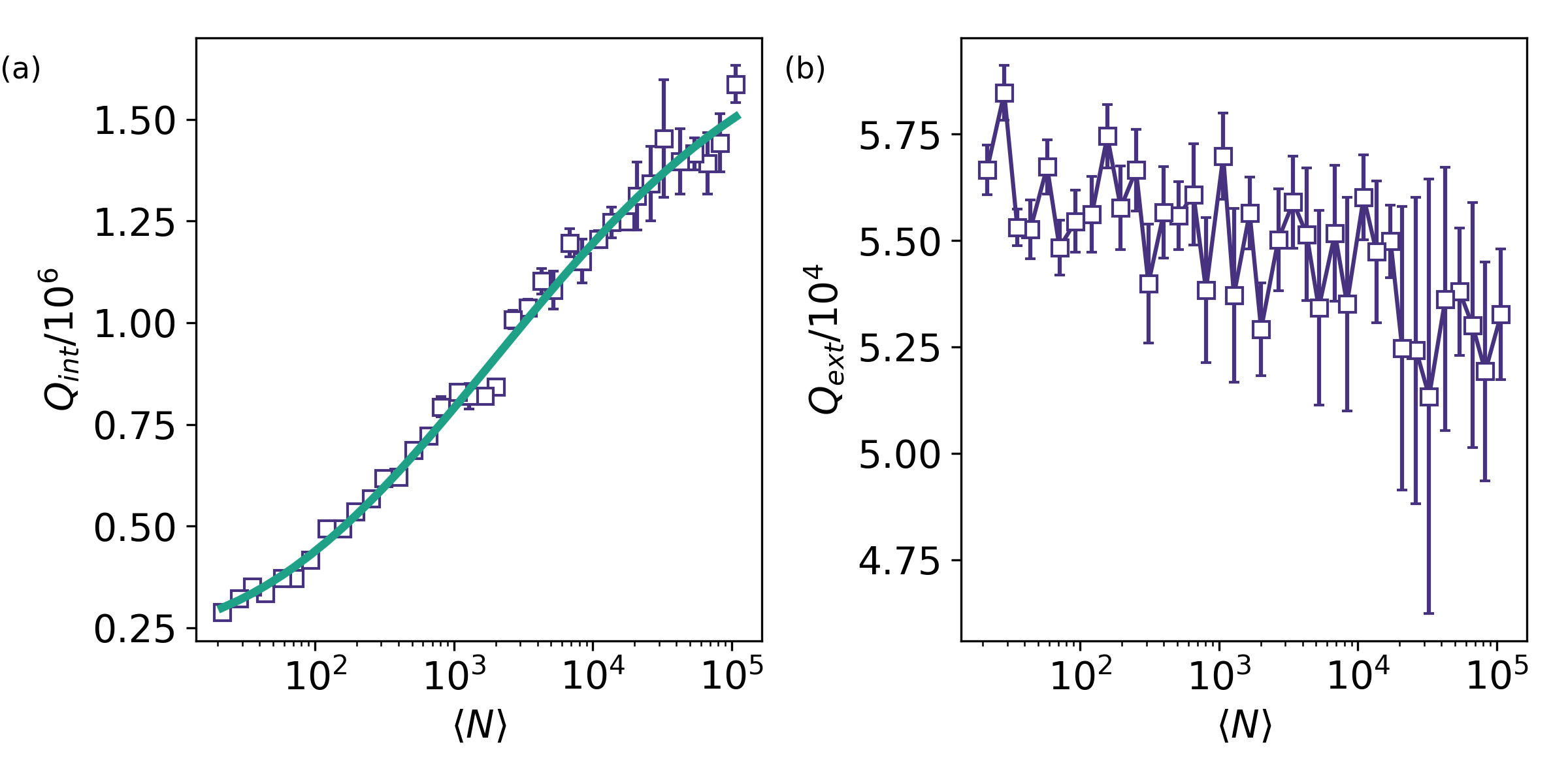}
\caption{\label{SI_fig:power-Q} \textbf{Quality factor of the resonator sample measured on the scanning gate microscope sample platform.} (a) Internal quality factor at different average photon occupancy. The solid line is a fit to the standard tunneling model (see text). (b) The external quality factor as a function of average photon number. Data taken at a temperature of 45 mK. Error bars are 95\% confidence bounds from fits.}
\end{figure}
\begin{figure*}[ht!]
\centering
\includegraphics[width=0.7\textwidth]{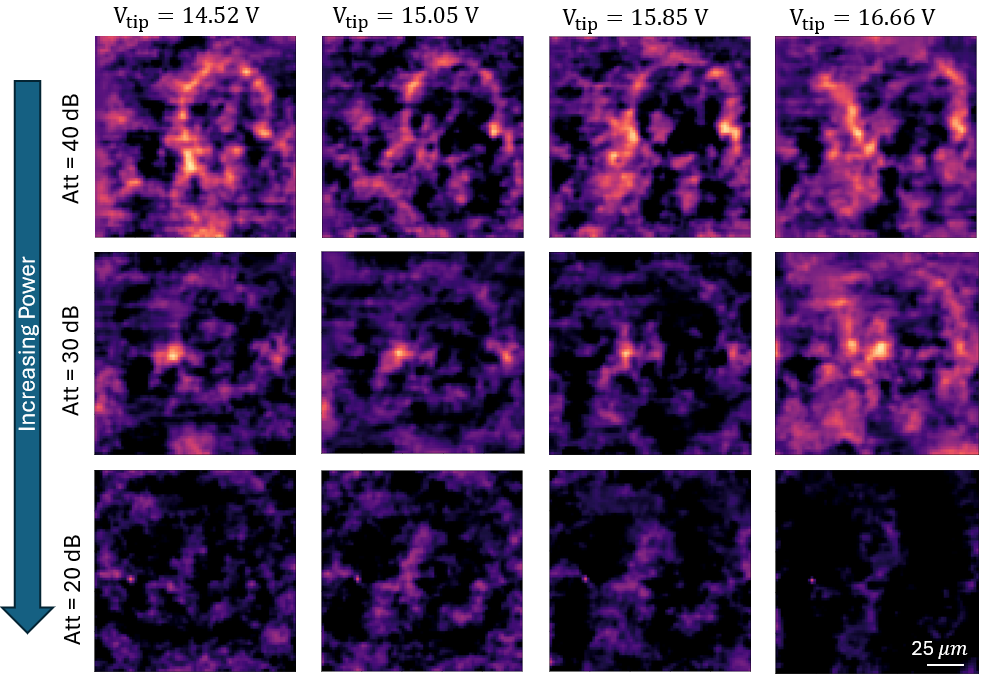}
\caption{\label{SI_fig:Pow} {\bf{Power dependence of TLS.}} Grids taken at the exact same location at low power (top panels) show more more pronounced TLS contours than those at larger powers (bottom panels). All panels are plotted on the same colour scale and a fitted background has been subtracted to highlight the fluctuations. All three grids were taken with the tip $15 \ \mu $m above the sample. The scale bar in the bottom right is shared across all panels. }
\end{figure*}

We estimate the average photon number by $\langle N \rangle  = Q^2 P_{\rm in}/(8\pi Q_{\rm ext} \hbar f_0^2)$, where $Q$ is the total and $Q_{\rm ext}$ the external (coupling) quality factors obtained from fits to the $S_{21}$ VNA data, and $P_{\rm in}$ is the microwave power reaching the sample. In Fig. \ref{SI_fig:power-Q} we show the internal and external quality factor as a function of $\langle N \rangle$. 
The data is fitted to $Q_{\rm int}^{-1} = F\tan\delta/(1 + \langle N\rangle /N_c )^\alpha + Q_{\rm int, 0}^{-1}$, finding $\alpha = 0.406\pm0.04$, critical photon number $N_c=13\pm 7$, TLS limited loss $F\tan\delta = (4.1\pm0.6)\times 10^{-6}$ and power-independent loss of $Q_{\rm int, 0} = (1.8\pm0.2) \times 10^6$. The quoted error bounds include propagated errors from the $Q_{\rm int}$ data. This strong dependence of $Q_{\rm int}$ on power indicates that the quality factor is strongly limited by TLS, which are saturated at increased driving powers. 

We further confirm TLS saturation by imaging individual TLS in our SGM setup. This is shown in Supplementary Fig. \ref{SI_fig:Pow}. For our experiments, we found that an average photon number in the range of $\langle N \rangle \approx 100-1000$ provides a good compromise between TLS sensitivity and signal-to-noise ratio. 

Supplementary Fig. \ref{SI_fig:Pow}, shows grids taken at the exact same location at different driving powers, by adjusting the variable attenuator on the signal input line shown in Supplementary Fig. \ref{SI_fig:RFsetup}.  From each panel, a background has been subtracted and all are plotted in the same colour scale. At low powers (high attenuation, top panels), we see a ring that grows with increasing tip voltage. The fluctuations reduce and ultimately disappear as the driving power is increased (bottom panels), showing that this individual TLS is saturated.

\section{Electrostatics modelling}
\begin{figure}
\centering
\includegraphics[width=0.4\textwidth]{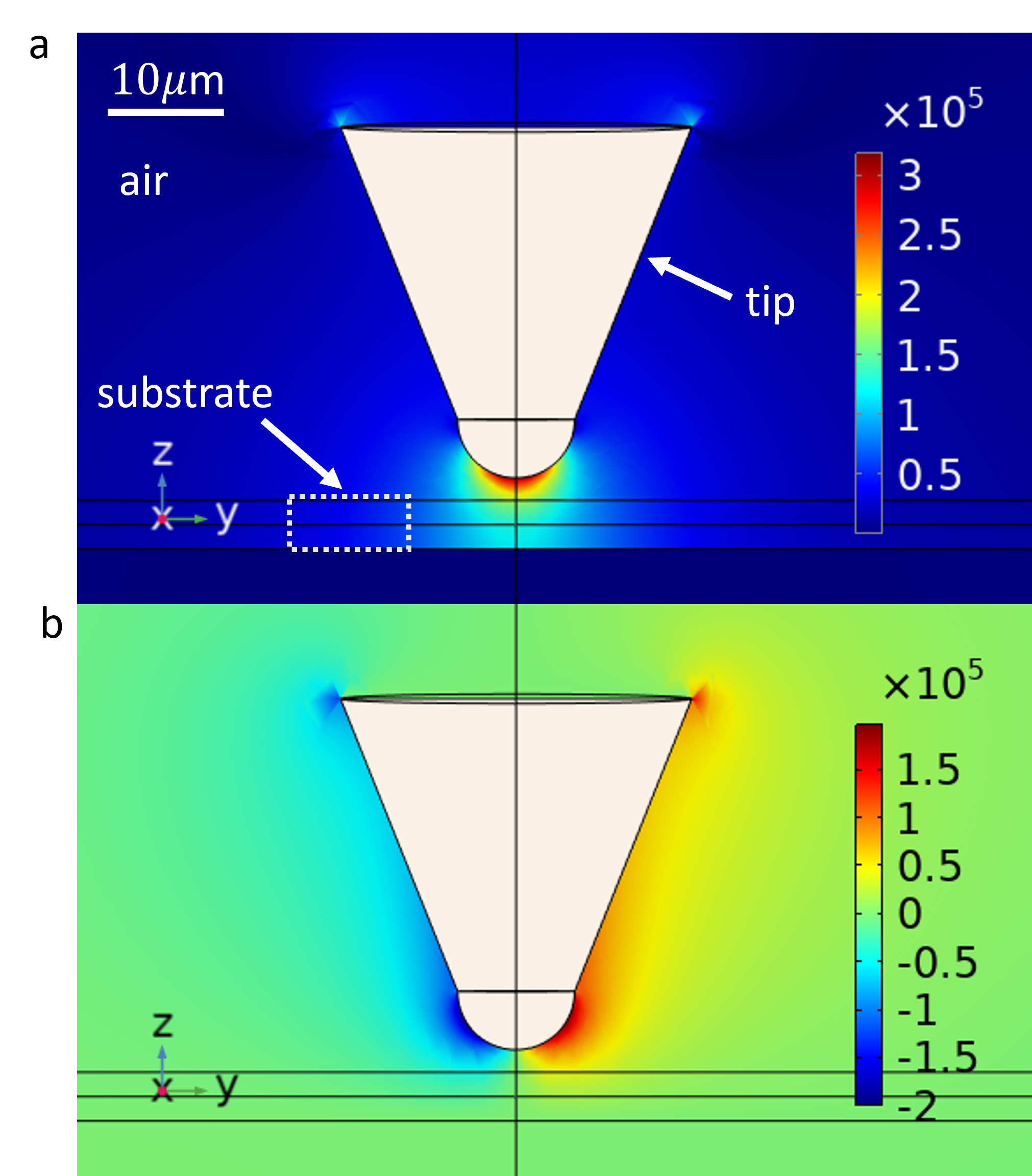}
\caption{\label{SI_fig:Tip Electric Field} \textbf{Simulated tip electric field.} (a) Tip Electric field in the z-direction $\abs{E_{z}}$, (b) Electric field in the y-direction. Color scale is electric field strength in units of V/m.
}
% Generated using 
\end{figure}

To simulate the frequency shift of a TLS and the resultant change in the $S_{21}$ transmission as it becomes resonant with the resonator, a simplified tip and sample geometry was simulated. In particular, a conical tip with a hemispherical bottom was chosen for the tip geometry (Supplementary Fig. \ref{SI_fig:Tip Electric Field}). The dimensions of the tip were chosen to be comparable to that of the actual tip dimensions, as measured by SEM imaging (in Supplementary Fig. \ref{SI_fig:tip}). A square $
100 \ \mu$m wide  and  $4 \ \mu $m thick substrate slab of sapphire under the tip imitated the sample. The underside of the sapphire slab was grounded and the tip was held at a potential of 1 V.   While this is a significant oversimplification of the sample geometry, the agreement between experimental and simulated results shows that our technique is able to capture the TLS dipole orientation even in the simplest of scenarios. A number of unknown experimental parameters (e.g. exact tip size and shape, exact local electric field strength, TLS location within the substrate/surface) will influence the exact determination of $\theta$. Future studies will undoubtedly have increased knowledge of these parameters. 

\begin{figure*}[ht!]
\centering
\includegraphics[width=1\textwidth]{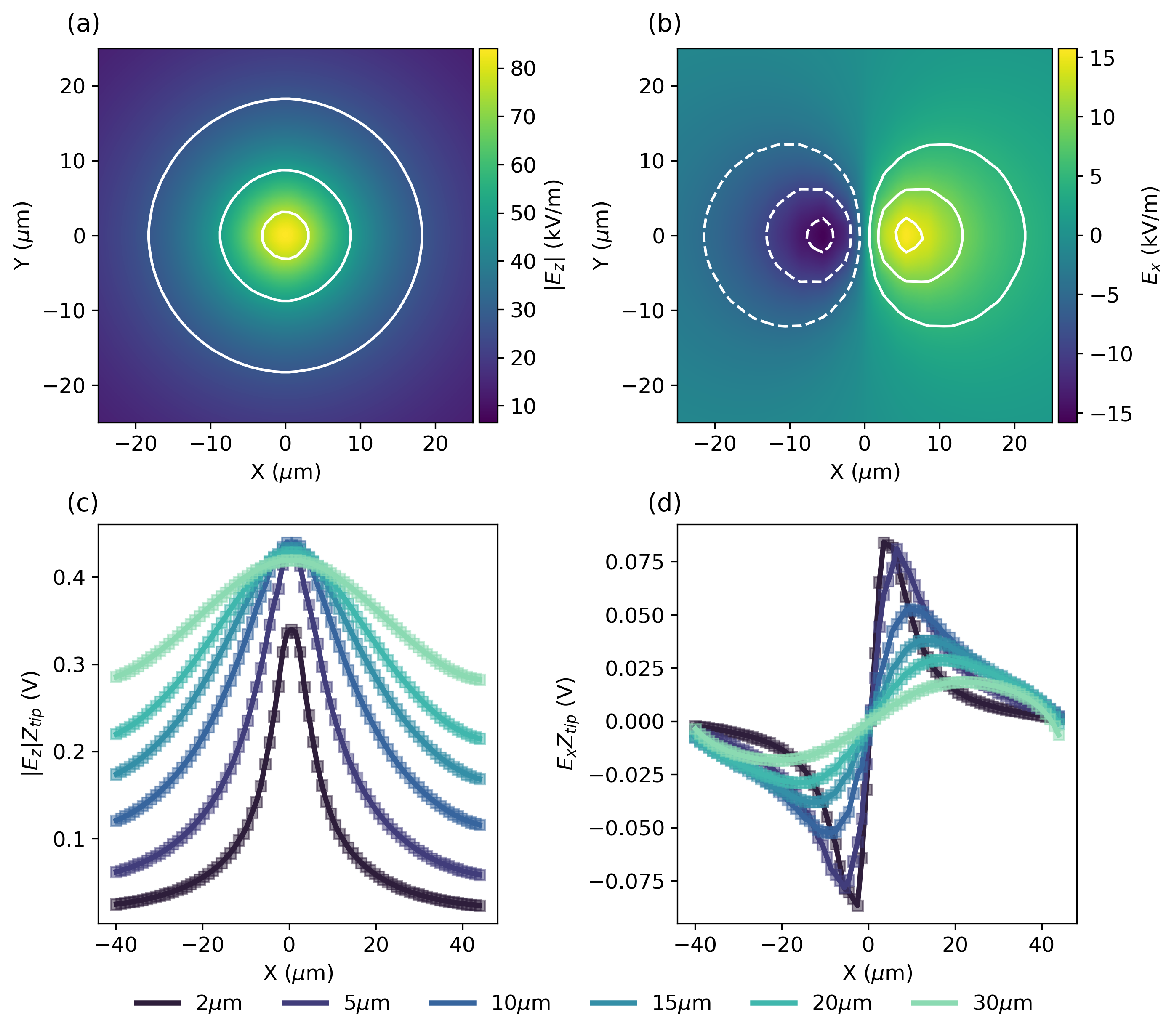}
\caption{\label{SI_fig:simulated-field-figures} \textbf{{Simulated tip electric fields at the sample plane.}} (a-b) Simulated electric field strengths in the sample plane for $Z_{\rm tip} = 5$ $\mu$m. (c-d) Cross-sections of simulated data similar to (a) and (b) for different $Z_{\rm tip}$ (different colors), normalised by the tip-sample distance $1/Z_{\rm tip}$.}
\end{figure*}

Supplementary Fig. \ref{SI_fig:Tip Electric Field} shows the magnitude of the simulated electric fields $\vec{E}_{DC}$ in the vertical ($E_z$) and horizontal ($E_y$) directions around the tip.

Similarly, Supplementary Fig. \ref{SI_fig:simulated-field-figures} (a-b) shows the resulting electric field strength in the sample plane, separated into the Z and X components respectively.

Supplementary Fig. \ref{SI_fig:simulated-field-figures} (c-d) compares the same electric field strengths as in Supplementary Fig. \ref{SI_fig:simulated-field-figures} (a-b) at $Y=0$ for different tip-sample separations for (c) $E_z$ and (d) $E_x$. 
Here we have scaled the data by the expected $E\propto 1/Z$ scaling. For $E_z$ we see that this almost collapses the curves at $x=0$ (some deviation due to finite tip size), and the larger $Z_{\rm tip}$ results in a more delocalised electric field distribution, as expected. In Supplementary Fig. \ref{SI_fig:simulated-field-figures}d we show the behaviour of $E_x$ with the same scaling applied. Here we see similar broadening, and we also see that the lateral component of the electric field vanishes much faster with increased tip-sample distance. I.e. ellipses could only be observed for small $Z_{\rm tip}$.

To mimic the change in voltage on the tip in the experiment we multiply the resulting $E_{DC}$ with a prefactor, before calculating the measured signal quantity through Eq. (2). Varying the tip voltage in simulation reproduces the change in size of the rings. As an example, in Supplementary Fig. \ref{SI_fig:R_vs_V} we plot the ring radius as a function of tip voltage, using parameters for the TLS resulting in a ring similar to that in Fig. 2 of the main text. Also, in simulations the ring radius shrinks approximately linearly with applied tip voltage, in analogy with Fig. 2h.

\begin{figure}[h!]
\centering
\includegraphics[width=0.35\textwidth]{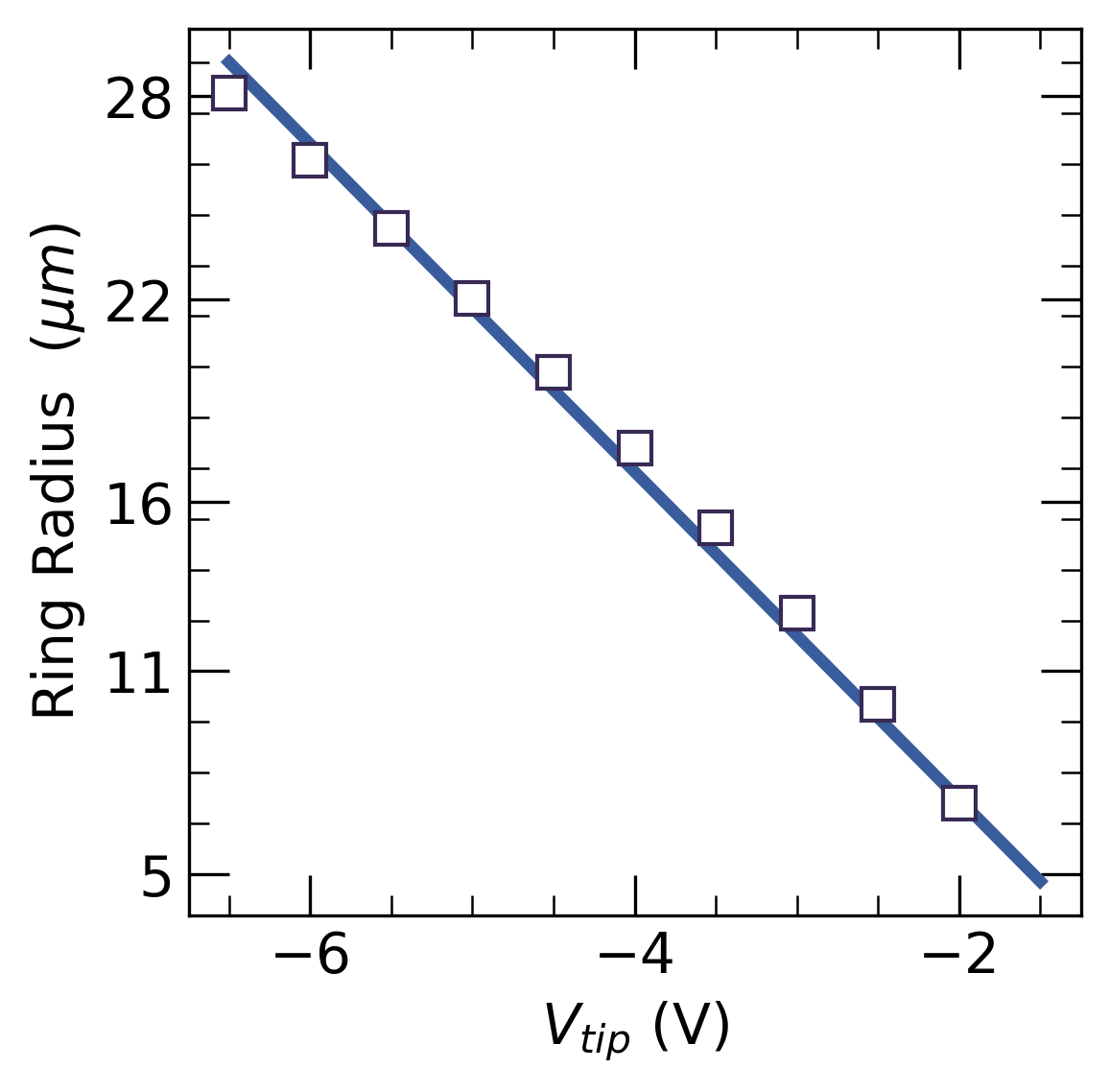}
\caption{\label{SI_fig:R_vs_V} \textbf{Simulated ring radius as a function of applied tip voltage.} A linear fit (solid line) to the ring radii for varying simulated tip voltages (markers). A dipole orientation of $\theta, \phi = 0$ was used. There is a clear linear dependence of the radius with voltage, as observed experimentally.}
\end{figure}
\clearpage

\end{document}